\documentclass[]{article}
\usepackage[english]{babel}
\usepackage{graphicx,enumerate}
\usepackage[tbtags]{amsmath}

\begin{document}
\title{Information and the arrow of time}
\author{Marcin Ostrowski}
\maketitle

\begin{abstract}
This paper is a discussion about the relationship between time and information. We argue that the direction of arrow of time is related to
the directivity of information copying that occurs in Nature.
\end{abstract}

\section{Introduction}

Why does time run forward? This question, on the borderline of Physics and Philosophy, has bothered mankind for centuries. According to some
people, this is not a valid question, since the impression of ``flow of time'' can only be a subjective sensation of our consciousness. The aim
of this work is not to determine the validity of this thesis. However, suspecting that there may be something to it, we will seek an answer to
another question:

\begin{quote}
{\it
Why do we remember the past and can not remember the future?
}
\end{quote}

\noindent
There must be some physical reason for this asymmetry. Even if the mere passage of time is subjective, such an asymmetry should not have a purely
subjective nature, but must be based on the laws of Physics (Nature). Otherwise, living organisms like ourselves, certainly would like to use
the knowledge about the future, since it is of strategic importance for survival. A thirsty animal would simply go to water using a certain
``memory of the future'', rather than head for the watering hole, remembering where it was yesterday, and having hope that today it is in the same
place (which may not always be true, for example, during intense drought). 
Another asymmetry between the past and the future, can be expressed by the question:

\begin{quote}
{\it
Why can we influence the course of future events and we can not influence the course of events from the past?
}
\end{quote}

\noindent
Generally, it is believed that the reason for this asymmetry is the second law of thermodynamics, but the problem does not seem to be settled
entirely. Let us quote here a chapter from a book by Rudolf Peierls\cite{peierls}:

\begin{quote}
{\it
``The ``arrow of time'' appears to be in our minds. As long as we have no clear explanation for this limitation, we might speculate whether
the time direction is necessarily universal, or wheter we could imagine intelligent beings whose time runs opposite to ours so that, from our
point of view, they could remember the future and make plans for the past.''
}
\end{quote}

\noindent
Our main aim will be to look closer at this issue.\footnote{The problem of time is widely discussed in the literature on the basis of physics and philosophy (for example \cite{extra}-\cite{extra4}). Ideas somewhat similar to ours can be found in (\cite{alberts}).}

\section{The symmetry of time reversal and the mechanics}

The problem of finding asymmetry between the past and the future seems to be particularly dramatic if we look closely at the laws of mechanics.
Both the Newton equations and the equations of modern physics treat equally both directions of time.\footnote{There is an exception, though,
in the world of elementary particles, namely there is some asymmetry between past and future in weak interactions. We are talking about the
violation of time reversal symmetry (T symmetry) during kaons decay\cite{kaony}-\cite{kaony2}. However, it seems that in Earth's conditions this phenomenon
is too marginal to be able to lead to such a profound asymmetry between the past and the future.}

\noindent
What follows from the symmetry of the Newton's equations? Consider, for example, the game of billiards.  Suppose that in an airtight container
we have $N$ particles perfectly elastically bouncing off the walls and from each other (as in the ideal gas model). Imagine that at a certain
fixed time $t_0$ we know the state of the system (i.e. the exact position and velocity of all particles in the case of a classic system).
We want to answer the following question: what will be the state of the system (gas), say in five minutes, i.e. at the time $t_0$+5min?
It turns out that the answer to this question is just as easy (or difficult) as the answer to the question: what was the state of the gas
five minutes earlier, i.e. at the time $t_0$-5min. Predicting the future is thus as difficult as predicting the past. So we have a full symmetry!

\noindent
One can give more examples of this kind. Take a relatively simple mechanical system like the Solar System. When in 2011 we ask: what will be
the position of planets in the sky on June 1, 3011, thus the answer to this question is as simple as the answer (using an appropriate computer
programme) to how it looked on June 1, 1011.

\section{Symbols of the past}

The fundamental difference between the past and the future is the occurrence of the memory of past events. These are the memories stored in
our minds, but also the contents of memory of non-living facilities such as computers. Computer memory, like ours, keeps the information recorded
in the past. So the inability to remember the future is not only an ailment of our brains but also brains of animals and memory of electronic
equipment.

\noindent
What is memory? It is not only the memory of humans, animals or computers. Also in the inanimate Nature there are processes that can be equated
with the memory. We can speak about memory when it is possible to read from a process (a phenomenon or an object) the information about another
process (a phenomenon or object). Examples of this are: the trace of rain imprinted on the rock millions of years before, fossil bones of animals
such as dinosaurs, the starlight emitted thousands of years earlier, finally, historical memorabilia: photographs, drawings, chronicles, etc.
Such objects will be called ``symbols of the past''\footnote{This term is coined by the author solely for the purpose of this work}.
It happens that in Nature we are confronted with symbols that ``speak'' of past, but we never meet similar symbols that ``speak'' of the future.
On the rock we never see traces of the rain which will be raining in some time, we can not see pictures that will be taken in some time, and we do
not observe light from stars that after some time will be emitted.
Of course, the past is not always ``well documented''. Most past events do not leave behind ``symbols of the past'' that are available for us.
Most of the rainfall after several hours does not leave a trace, as well as most of the events are not photographed. This is especially a problem
for historians studying ancient times, archaeologists and palaeontologists.

\section{Are there any ``symbols of the future''?}

The future does not offer us memories. Does this mean that, unlike the past, we do not have any knowledge about future events? Do we have something we might call ``symbols of the future''? The answer to this question is yes. ``Symbols of the future'' could relate to our projections or plans for the future.
For example, when we organise a trip or a holiday, we can schedule when we will get out of home, how we will get to the station, which train we will catch, what we will visit every day. Those plans, made by us a priori, are ``symbols of the future''.  With a bit of good organisation and luck, we
can put them into practice. It is true that we do not always succeed, because we can not predict everything entirely. However, if our live is ordered, then it often becomes exceedingly predictable (sometimes to our bitterness).

\noindent
There is, however, a crucial difference. In the case of ``symbols of the past'' we just remember what has already happened, whereas in the case of ``symbols of the future'' we only provide or plan, what will happen. But we forget for a moment about the difference between memories (symbols of the past) and forecasts (symbols of the future), treating them as if the difference in their perception was only a subjective impression of our brain.

\noindent
However, ``symbols of the future'' are not certain. The plans that we make (or our predictions) often do not work. But the problem is that ``symbols of the past'' are not certain either. Their number may be insufficient or sometimes they can be false. For example, if the offender intentionally dissects false evidence to confuse the trail of the investigation.

\noindent
Are we back to square one then? Uncertain ``symbols of the past'' do not guarantee that we remember the past well and uncertain ``symbols of the future'' make it impossible to predict the future exactly.

\section{Redundancy of the ``symbols of the past''}

However, there are objective differences between ``symbols of the past'' and ``symbols of the future''. We could say that the uncertainty of both
have a completely ``different calibre''. In the case of ``symbols of the future'', it is often enough to change one of them to make the entire
prediction fail. As an example, re-consider the planning of a holiday travel. We choose a place and time, reserve the accommodation, check rail
connection, buy a ticket in advance, inform friends and take a leave. All of these things (i.e. reservations stored in computer memory or in a printed
form, record in our minds and the minds of our friends) are examples of ``symbols of the future''. But when one of them collapses, for example if we
do not get a leave within the planned period because in the last minute some important issues come out, the whole trip will not take place. The fact
that other symbols still support our intention is not relevant here.

\noindent
In the case of the ``symbols of the past'', the situation is different. They are redundant to each other and erasing (or misrepresenting), one
of them does not change  the authenticity of the event. For example, imagine a man who was a participant of a trip to Paris
last year. How can we be confident (say, after one year) that this man actually was on this trip? This is evidenced (among others) by: his memory,
the photos that he made using his camera and photos taken by others, showing him against a background of the monuments of Paris. All these symbols
affirm each other and there is redundancy in them. Imagine that during this trip a crime was committed (say, theft) and it becomes of interest to
the police. Imagine that the person in question became a suspect in this case. Imagine also that after testimonies of all the participants of the
trip and after analysis of pictures from their cameras, the following situation emerged: half of the participants said that this man was actually
on the trip, while the other half strongly denies it. The man in question appears in half of the pictures, but he does not appear in the other half.
Moreover, the police reported another man who exactly at the same time, was on another trip, say, in Rome, claiming that the suspect was there with
him. Again, half of the participants (and their photos) confirms that the suspected participation in the trip, while the other half denies it.
What would be the surprise of the detectives if the facts turned out that way. Fortunately, physical reality does not play such tricks on us. Even
if there were some inconsistencies in the testimonies of members of both tours, we would be more inclined to believe in a deliberate lie of one group
aimed at providing the suspect with an alibi than in any fluctuations in reality.

\noindent
Thus, it is much easier to change the future by modifying ``symbols of the future'' than change the past by introducing  modifications to ``symbols
of the past''. Just a train accident or a strike can hold the planned trip. In the case of a past travel, it is not enough to falsify or deface one
picture. This should be done with all the photos of all participants, with a record of memories of their brains, and many symbols contained in the
environment.  It is highly unlikely or impossible. Let us summarise:

\begin{quote}
{\it
``Symbols of the past'' in a redundant manner mutually confirm one version of past events. In the case of the ``symbols of the future'' a change
of even one of them often alters the course of future events. 
}
\end{quote}

\section{Symbols of the past as copies of the information}

What are ``symbols of the past''? To answer this question, we refer here to a discussion of our earlier work \cite{moja}. We identify there the
information with the abstract quantity that undergoes copying. What are ``symbols of the past'' if not just copies of the information about the
events of the past?

\begin{quote}
{\it
``Symbols of the past'' are copies of information about events of the past.
}
\end{quote}

\noindent
There may be many symbols of one event, for example when we take a photograph of an object (e.g. the Eiffel Tower). On a sunny day, billions of photons are reflected from the Eiffel Tower every second. Thus, potentially, by registering the photons, it is possible to take millions of photos, i.e. copies of information about the object. Even if we do not do this, there will be information contained in the photons escaping into outer space all the time. So we have a huge redundancy. We mean by it (firstly) that symbols of the past (copies of information) can be numerous, and (secondly) that they are mutually non-contradictory.\footnote{We observe it especially when we read different books on the same subject. Each of them brings less and less to the case, until, finally, in the next one, most of the facts are duplicates of what we have read
elsewhere (maybe with the exception of a few paragraphs)} This redundancy occurred in ``symbols of the past'' leads to significant differences in the way we perceive the future and the past.

\noindent
So why do we remember the past? Because it is sufficient to keep in mind one of the many copies of information about the incident to know that it
happened. Then we do not need to know the state of the entire system (as in billiards), to be able to judge about its past. Between ``symbols of
the past'' we have a strong correlation (redundancy). It is sufficient to know only a few of them (e.g. holiday photos), and the rest will overlap
with them, not contributing much to the case. Therefore, despite the fact that comprehensive information about events that happened a year ago is
located in an area with a radius of 1 light year, thanks to the redundancy such comprehensive information is not often needed. Sometimes, it is enough
to know a small percentage of all ``symbols of the past'' in order to know what happened.

\noindent
For the future, it is not the case. We must know the state of the entire system to be able to predict anything about its future. To know what will
happen exactly in a year, you need to know the state of all particles from the area with a radius of 1 light year. Each individual change may yet
spoil everything (change the course of history).

\begin{quote}
{\it
The difference between the past and the future seems to stem from the direction of copying information. For some reason the information about
past events is copied into the future, not vice versa.
}
\end{quote}

\noindent
Symbols of the past, as we said earlier, are copies of information about past events. But most importantly, they are easily readable copies. What
do we mean by that? Referring to the unitarity of time evolution in Quantum Mechanics, we can conclude that if we burn a book, then the smoke and
ash formed after combustion will include the entire information that was contained in it. The problem is that there does not exist (and probably will
not exist) a technical method of reconstructing this information. Thus, burning documents is still the best way to get rid of inconvenient evidence.
The situation is different in the case of copies that are easy to read. Light reflected from the printed sheet carries the information about its contents
and is an example of a copy easy to read. Why? Because we only need the lens and the CCD to reproduce from a part of the reflected light the form of
printed text. This is what we call the ``copies easy to read''. For some reason the past offers them for us, and the future does not. Unlike ``symbols
of the past'', ``symbols of the future'' are not copies of the information about future events, but they are just our assumptions, intentions and
sometimes only dreams.

\noindent
When we remember the past we are not using unitarity of time evolution. If we did, then we could read a book, watching the smoke emitted from burning
it. What human brain does is register easy to read ``symbols of the past'' and memorise them by becoming their database. Most importantly, the way
the symbols are stored (i.e., memory organisation) is that they are still easy to read. Memory of the past does not lie in capturing the current state
of the particles and simulating where they must have been in the past. The brain is not able to do that and never will be (it would have to have
something of a demon as the Maxwell's demon). If that were the case, then remembering the past and predicting the future would be equally easy for it.
Then, for the brain, the past would not be different from the future.

\noindent
Let us now turn to the second question: ``Why can we change the future and we can not change the past?'' We can not change events from the past because
we will never erase all copies of the information about it. Especially when those copies are numerous. For the future, the situation is different.
We do not need to remove multiple copies of information about future events because they do not yet exist. The result is that it is much easier to
modify the future than the past.

\section{Irreversibility}

Copying of information is a reversible process.\footnote{We can illustrate it by the simplest example. Consider two lines of numbers. In the first
line we have: ``01011100011'' while in the second line there are only zeros. The first line is treated as data that we want to copy, the second is
a medium prepared for copying. Zeros mean that the medium is clean as a white sheet of paper ready to write on. Copying will result in a situation
where in both lines there will be the same string: ``01011100011''. Is it possible to reverse this copying? That is, is it possible to reconstruct
the situation prior to copying? Of course it is possible. It is enough to write only zeroes in the bottom line. The case would look differently if
we did not have a clean medium in the form of lines of zeros. Then its preparation would require erasing the original recording (noise), and therefore
it would be irreversible.}
Therefore, we should be able to erase the past, if we can reverse the copying processes and delete all copies in which information about an event is
recorded.\footnote{It's a bit like in the movies crime: perfect crime when is possible to dispose of all witnesses of crime.}
These types of experiments can be performed in case of simple systems (i.e. consisting of a small number of particles). They are the so called quantum
erasure experiments \cite{lit1}-\cite{lit2}. They are based on the reversibility of unitary evolution of an isolated quantum mechanical system. The problem emerges
when some witnesses of the event (particles carrying the information about it) ``escape from the crime scene''. Then the erasure of the event is not
possible. This is a particular problem in the case of photons. They have an ``unpleasant'' habit of escaping with the highest speed in nature. If such a
photon escapes into the empty space (vacuum) nothing is able to overtake it and force it to turn back (unless during earlier preparation for the
experiment, a sufficient number of mirrors have been set). When an event happens, for example, on the surface of the Earth on a clear day, the photons
reflected every second from our body have virtually no barriers to fly away into space. Each of them carries a small piece of information about us
(about our position, appearance, etc.). It is true that each of them is running in a slightly different direction, and carries in itself too little
useful information. However, would it not be possible in this situation to focus the photons and collect back the information carried by them?

\noindent
To answer this question let us look at the equation of resolution of light microscope:
\begin{equation}
d=\frac{\lambda}{\sin\alpha}.
\end{equation}
The most important moral of Eq.~(1) is that the resolution $d$ of such a microscope is of the order of a wavelength $\lambda$ which is emitted
by the observed object (specimen). For this purpose, the aperture angle $\alpha$ must be of the order of $90^0$ i.e. the photons reflected from
an object must be collected by a lens with the widest possible range of angles. But that is not what we mean here. The most important thing we
should note is that in the formula (1) there does not exist the distance between the microscope lens and the observed object. This distance does
not affect the resolution in any way. This means that regardless of whether it is of the order of millimetres or light years, resolution that we
get is the same (we omit the possible opacity of the medium which at large distances can introduce significant distortions). Obviously, for a large
distance between an object and a lens, in order to obtain a sufficient angle of aperture, the lens needs to be sufficiently large. However, this is basically a technical problem.\footnote{For example, if we take the resolution $d=1$ mm observing at wavelengths $\lambda=500$nm, then we have from Eq.~(1) $\sin\alpha=5*10^{-4}$, which gives for a distance of observation equal to 100 light-year lens radius equal to 0.05 light-years. A suitable gravitational lens e.g. in the form of properly shaped transparent dust giving an appropriately shaped gravity field should work.}

\noindent
We certainly do not claim that this type of experiment is feasible. This should be regarded more as a thought experiment, showing that information
about events from distant past, which did not leave a trace on Earth, can exist in space and, theoretically, can be read. Thus, those events can not
be erased (reversed). Surely, this can not be done locally, being just on Earth. Therefore, such an event appears irreversible.\footnote{Obviously
we can not assume in advance that we are not part of some grand time erasure experiment, run on a cosmic scale, in which a large number of properly
arranged mirrors changes the direction of movement of all the photons, and the event is erased. However, this seems quite unlikely.}

\section{A question about the essence of time}

What have we been able to do? We have replaced two questions of a rather subjective form (``Why do we remember the past and do not remember the
future?'' and ``Why can we only change the future?'') with one:

\begin{quote}
{\it
Why is the information copied only in one direction in time?
}
\end{quote}
The advantage of a question posed in such a way (in comparison with the previous two) is that it has a physical nature and is not related to fields
such as philosophy or psychology.

\noindent
At this point we could actually finish the discussion, because the answer to this question is already a case for a completely different story.
But we will allow ourselves to speculate and present a some hypothesis.

\noindent
Making multiple copies of information is possible, if we have sufficient number of free media to record it. In our world, these media are particles (atoms, electrons,
photons) and their conglomerates (condensed phase). However, the total number of particles in the Universe increases over time. In particular, this applies to photons.\footnote{In the case of other particles
such as electrons or protons the situation is different. Their existence is subject to principles of conservation (of charge, of lepton number, of baryon number).}
Photons are very important, as they arrive from the Sun to the Earth, drive the majority of processes on the
planet and do not allow multiple subsystems to achieve the thermodynamic equilibrium (this applies particularly to biological subsystems). Where do
these photons arise? They are produced by thermonuclear fusion reactions occurring in the core of the Sun. The merger of four protons in helium nucleus
forms a few gamma-ray photons with the energy of the order of MeV. However, before these photons reach the photosphere, they are repeatedly,
inelastically scattered, dividing into a the number of secondary photons with lower energy. The surface of the Sun (photosphere) emits photons of
average energy of the order 2eV (equivalent to yellow light). The overall balance of gains and losses for particles associated with the act of producing
one helium nucleus is as follows\footnote{What may arouse some concern is the presence of two electrons on the substrate. Of course they are not
directly involved in the cycle p-p, but they annihilate with two positrons produced in the first reaction of the cycle.}:

\begin{equation}
4\,^1_1H+2\,e^-\rightarrow\,^4_2He+2\nu_e+10^7\,\gamma
\end{equation}

\noindent
In a world where there is a $10^7$ additional photons (right side of Eq.~2) there are much more media to store many of the new symbols of the past,
while it is difficult to imagine this in a world in which there are six particles (left side of Eq.~2). It is natural that in states with a greater
number of particles there will appear the information (symbols of the past) about states where there were fewer such particles, but not the other
way round. Then we identify in a natural way states with a small number of particles as a remembered past, and states with more particles as the
unknown future. So we can say that the future embraces the past, while the past does not embrace the future. Thus, the moment which corresponds to
the situation on the left side of the equation corresponds to the past, while the right to the future.

\noindent
One might ask, ``Why did the Universe begin with the state of the (relatively) small number of particles having a high density and energy?''
But this question is ill-posed. We only need to ask why such a state was among the states which were assumed by the Universe. Because if such
a state is one of the possible states of the Universe, it naturally will be the beginning of time.

\noindent
If we identify (naively) entropy $S$ with the number of particles in the Universe, this might justify the conclusion that it is not the entropy
that increases as a result of the passage of time, but conversely, the subjective feeling of time passage follows the direction of entropy increase.

\section{Chaos and the direction of the arrow of time}

Sometimes it is suggested that the direction of the arrow of time can be associated with the existence of chaos \cite{nowy1}. It is also claimed that the transition from the description based on trajectories to the statistical one (based on a probability distribution) guarantees the distinction  of on directions of time. For example, according to \cite{nowy1}, random walking describes an unidirectional increase of blurring of the particle's position towards the future.

\noindent
In our opinion, it is exactly the opposite! Chaos not only does not determine the arrow of time, but in fact it blurs the distinction between the past and the future. If all systems in nature were chaotic, the difference between the past and future would be erased as a result of the destruction of all ''symbols of the past''.\footnote{
As a practical example, let us consider such ''symbols of the past'' as the fossils from the earlier geological epochs. It turns out that fossils older than 200million years are found only in the sediments underlying the continental crust, but never in the oceanic crust \cite{nowy2}. This is caused by the participation of the oceanic crust in the convection of the Earth's mantle. The oceanic crust is formed constantly in the spreading zones and destroyed (melted) in the subduction zones. Thus, the age of the oldest parts of the oceanic crust is up to 200 million years.

\noindent
In the case of continental crust, the situation is different! It does not participate directly in the convection.  Its fragments are only moving only across the surface, but are not dragged into the Earth and melted. Thanks to this, in the sediments accumulated on its surface, it stores evidence of biological life which is much older than 200 million years. We can say that the continental crust is like a hard drive which stores data about the geological history of the planet.

\noindent
Well, but what does it all have to do with chaos? The answer is that convection is an example of a chaotic process \cite{nowy2}. If the entire Earth's crust was involved in this process, our understanding of life, for example, in the Cambrian, would be as scarce as our knowledge about the life on Earth in the next 500 million years.}

\noindent
We could take the point of view presented in \cite{nowy1} if we accepted the existence of objective probability of random events, associated with low entropy at the beginning of the Universe and then evolving in the direction of the entropy's growth. However, we assume the ''subjective'' interpretation \cite{moja}. We believe that the probability distribution is our knowledge about the current state of a random process. The increase of entropy with time means that our prediction about the future state becomes less accurate. However, we can make a prediction about the past in the same way. We can define the entropy decreasing in time, which means that our forecast of the state of the process for more distant past also becomes less accurate.

\noindent
The author of  \cite{nowy1} admits that in the mathematical equations describing a random processes there are components corresponding both to the increase and to the decrease of entropy over time. But the latter are cut arbitrarily as having no reference to reality. However, such an approach is an arbitrary imposition of one direction of time, and can not be considered as a proof of irreversibility of processes in nature.

\noindent
Why are weather forecasts not made for the past? The reason is not because the probability is asymmetric in time, but because such forecasts are unnecessary (there are ''symbols of the past''). If such forecasts were needed, they would be equally difficult to create and their results would be sometimes equally surprising as those for the future.\footnote{Do you remember the movie ''50 First Dates''? Just imagine what would happen if all humans (and other data carriers) had the same ailment as the main female character, and the rain puddles from the day before dried up very quickly (high speed of chaotic phenomena ).}

\noindent
Contrary to the tendencies advocated by some authors, neither the concept of probability nor the logic distinguish any direction of time. The direction of the arrow of time on Earth is determined mainly by the direction of the Sun's rays. It is an irreversible, but not a chaotic process.

\end{document}